\documentclass[iop, apjl]{emulateapj}
\usepackage{threeparttable}
\usepackage{multirow}
\usepackage{times}
\usepackage{enumitem}
\usepackage{url} 
\usepackage{amsmath,bm}
\usepackage{color,graphicx}

\shorttitle{}
\shortauthors{Y. Huang et al.}

\begin{document}

\title{Discovery of two new hypervelocity stars from the LAMOST spectroscopic surveys}

\author{Y. Huang\altaffilmark{1,2,7}}
\author{ X.-W. Liu\altaffilmark{1,2,3}}
\author{ H.-W. Zhang\altaffilmark{2,3}}
\author{ B.-Q. Chen\altaffilmark{1}}
\author{ M.-S. Xiang\altaffilmark{5,7}}
\author{ C. Wang\altaffilmark{2}}
\author{ H.-B. Yuan\altaffilmark{4}}
\author{ Z.-J. Tian\altaffilmark{2,7}}
\author{Y.-B. Li\altaffilmark{5}}
\author{B. Wang\altaffilmark{6}}
\altaffiltext{1}{South-Western Institute for Astronomy Research, Yunnan University, Kunming 650500, People's Republic of China; {\it yanghuang@pku.edu.cn {\rm (YH)}; x.liu@pku.edu.cn {\rm (XWL)}}}
\altaffiltext{2}{Department of Astronomy, Peking University, Beijing 100871, People's Republic of China {\it zhanghw@pku.edu.cn (\rm HWZ)}}
\altaffiltext{3}{Kavli Institute for Astronomy and Astrophysics, Peking University, Beijing 100871, People's Republic of China}
\altaffiltext{4}{Department of Astronomy, Beijing Normal University, Beijing 100875, People's Republic of China}
\altaffiltext{5}{Key Laboratory of Optical Astronomy, National Astronomical Observatories, Chinese Academy of Sciences, Beijing 100012, People's Republic of China}
\altaffiltext{6}{Key Laboratory for the Structure and Evolution of Celestial Objects, Yunnan Observatories, CAS, Kunming 650216, People's Republic of China}
\altaffiltext{7}{LAMOST Fellow}

\begin{abstract}
We report the discovery of two new unbound hypervelocity stars (HVSs) from the LAMOST spectroscopic surveys.
They are respectively a B2V type star of $\sim$\,7\,M$_{\rm \odot}$ with a Galactic rest-frame radial velocity of 502\,km\,s$^{-1}$ at a Galactocentric radius of $\sim$\,21\,kpc and a B7V type star of $\sim$\,$4$\,M$_{\rm \odot}$  with a Galactic rest-frame radial velocity of 408\,km\,s$^{-1}$ at a Galactocentric radius of $\sim$\,30\,kpc.
The origins of the two HVSs are not clear given their currently poorly measured proper motions.
However, the future data releases of Gaia should provide proper motion measurements accurate enough to solve this problem. 
The ongoing LAMOST spectroscopic surveys are expected to yield more HVSs to form a statistical sample, providing vital constraint on understanding the nature of HVSs and their ejection mechanisms.
\end{abstract}

\section{Introduction}
%the definition & mechanism 
Hypervelocity stars (HVSs) are rare objects with extreme fast velocities that can exceed the Galactic escape speed.
It has been suggested that they originate near the Galactic center (GC) by dynamical interactions between (binary) stars and the central massive black hole(s) (MBH; e.g. Hills 1988; Yu \& Tremaine 2003).
In addition to the GC origin, alternative models proposed to explain the HVSs include the tidal debris of an accreted and disrupted dwarf galaxy (Abadi et al. 2009), the surviving companion stars of Type Ia supernova (SNe Ia) explosions (Wang \& Han 2009), the result of dynamical interaction between multiple stars (e.g, Gvaramadze et al. 2009), and the runaways ejected from the Large Magellanic Cloud (LMC), assuming that the latter hosts a MBH (Boubert et al. 2016).

%Current observation
The first HVS, a B-type star with a Galactic rest-frame radial velocity of 673\,km\,s$^{-1}$ in the outer stellar halo, was discovered serendipitously by Brown et al. (2005).
Hitherto, over 20 HVSs have been found from both systematic searches (Brown et al. 2006, 2009, 2012, 2014; Zheng et al. 2014) and by serendipitous discoveries (Brown et al. 2005; Hirsch et al. 2005; Edelmann et al. 2005; Heber et al. 2008; Geier et al. 2015).
Most of them are massive stars (larger than $2$--$3$\,M$_{\odot}$) of spectral type B in the Galactic halo.
The observed properties (e.g. the spatial and velocity distributions) of these HVSs are consistent with a GC origin for ejecting unbound HVSs (Brown 2015).
The only two exceptions are the stars US\,708 and HD\,271791.
The former is a helium-rich subdwarf O-type star with a mass of $\sim$\,0.3\,M$_{\odot}$.
It is also the currently known fastest unbound star in our Galaxy with a total velocity of about 1200\,km\,s$^{-1}$, and its reconstructed trajectory disfavors a GC origin but favors a donor remnant of a helium double-detonation SNe Ia (Wang et al. 2013; Geier et al. 2015).
HD\,271791 is an 11\,M$_{\odot}$ B-type star and its proper motions suggest that it is an unbound runaway disk star ejected by binary interactions (Heber et al. 2008).

%importance
HVSs are powerful tracers to probe the mass distribution of the Galaxy since they travel large distances across the Galaxy (e.g. Gnedin et al. 2005; Kenyon et al. 2008).
More importantly, in the era of {\it Gaia}, accurate measurements of their three dimensional motions become possible.
The measurements provide constraints on the shape of the dark matter halo of the Galaxy.

%this work
As discussed by Zheng et al. (2014; hereafter Z14), the LAMOST spectroscopic surveys have the potential to yield a large statistical sample of HVSs.
They performed a systematic search for HVSs in the first internal Data Release (DR1) of LAMOST and discovered a B-type HVS with a Galactic rest-frame radial velocity of 477\,km\,s$^{-1}$, located at a Galactocentric radius $r$ of 19.4\,kpc.
By June 2016, LAMOST collected nearly 6.5 million stellar spectra of signal-to-noise-ratios (SNRs) better than 10.
The sample size is three times larger than the DR1 used by Z14.
In this letter, we continue the systematic search for HVSs using the latest data release of the LAMOST spectroscopic surveys.
In Section\,2, we briefly describe the data and target selection.
The results are presented  and discussed in Section\,3.
Finally, we summarize in Section\,4.

\begin{table*}
\centering
\caption{Properties of the three HVSs discovered by LAMOST}
\begin{threeparttable}
\begin{tabular}{llll}
\hline
&LAMOST-HVS1&LAMOST-HVS2&LAMOST-HVS3\\
\hline

{Position ($J2000$)}&($\alpha$, $\delta$) = (138\fdg027199, 9\fdg272725)&($\alpha$, $\delta$) = (245\fdg086520, 37\fdg794456)&($\alpha$, $\delta$) = (50\fdg321157, 19\fdg126719)\\
                                &($l$, $b$) = (221\fdg099543, 35\fdg407234)&($l$, $b$) = (60\fdg398889, 45\fdg253192)&($l$, $b$) = (165\fdg140088, -31\fdg200353)\\

Magnitudes&$B = 12.94 \pm 0.04$&$B = 14.98 \pm 0.03$&$B = 16.63 \pm 0.02$\tnote{a}\\
&$V = 13.06 \pm 0.01$&$V = 15.10 \pm 0.03$&$V = 16.69 \pm 0.02$\tnote{a}\\
&$J = 13.36 \pm 0.03$&$J = 15.49 \pm 0.05$&$J = 16.69 \pm 0.14$\\
&$H = 13.43 \pm 0.04$&$H = 15.52 \pm 0.11$&$H = 16.61 \pm 0.29$\\
&$K_{s} = 13.53 \pm 0.04$&$K_{s} = 15.57 \pm 0.21$&$K_{s} =15.88$\tnote{b}\\

$E(B-V)$&0.055&0.015&0.115\\

Heliocentric distance (kpc)\tnote{c}&$13.91 \pm 1.97$&$22.24 \pm 4.57$&$22.32 \pm 2.50$\\

Radial velocity  (km\,s$^{-1}$) &$v_{\rm los} = 611.65 \pm 4.63$&$v_{\rm los} = 341.10 \pm 7.79$&$v_{\rm los} = 361.38 \pm 12.52$\\
                                                &$v_{\rm rf} = 473.52 \pm 5.35$&$v_{\rm rf} = 502.33 \pm 8.37$&$v_{\rm rf} = 408.33 \pm 12.57$\\

Spectral type&B1IV/V&B2V&B7V\\

Absolute magnitude&$M_{V} = -3.10 \pm 0.50$&$M_{V} = -1.70 \pm 0.63$&$M_{V} = -0.40 \pm 0.33$\\
&$M_{J} = -2.43 \pm 0.50$&$M_{J} = -1.24 \pm 0.63$&$M_{J} = -0.16 \pm 0.33$\\
&$M_{K_{s}} = -2.33 \pm 0.50$&$M_{K_{s}} = -1.10 \pm 0.63$&$M_{K_{s}} = -0.08 \pm 0.33$\\

Flight time (Myr)\tnote{d}&$38 \pm 4$&$40 \pm 8$&$62 \pm 6$\\ 

Mass ($M_{\odot}$)&11.0&7.3&3.9\\

Luminosity ($L_{\odot}$)&13490&2399&309\\

$T_{\rm eff}$ (K)&26000&20600&14000\\

Life time (Myr)&$\sim$\,$10$&$\sim$\,$16$&$\sim$\,$80$\\

$r$ (kpc)&$20.14 \pm 1.98$&$20.86 \pm 4.57$ &$29.59 \pm 2.51$\\

$Z$ (kpc) &$8.06 \pm 1.14$&$15.80 \pm 3.25$&$-11.56 \pm 1.30$\\

Note & Zheng et al. (2014) & This work & This work\\

\hline
\end{tabular}
\begin{tablenotes}
\item[a] The Johnson $B$ and $V$ magnitudes are converted from the $g$ and $r$ magnitudes measured by the SDSS \& XSTP-GAC survey.
\item[b] Lower limit with a large uncertainty.
\item[c] The heliocentric distances are adopted to be the weighted mean values derived from $V$, $J$ and $K_{s}$ bands if they are available.
\item[d] The flight time means the time required for an HVS to travel from the GC to its current position. 
%Here we note the flight times derived here are actually upper limits since we assumed that the observed radial velocities are the full space motion of the HVSs. 
\end{tablenotes}
\end{threeparttable}
\end{table*}

\begin{figure*}
\begin{center}
\includegraphics[scale=0.4,angle=0]{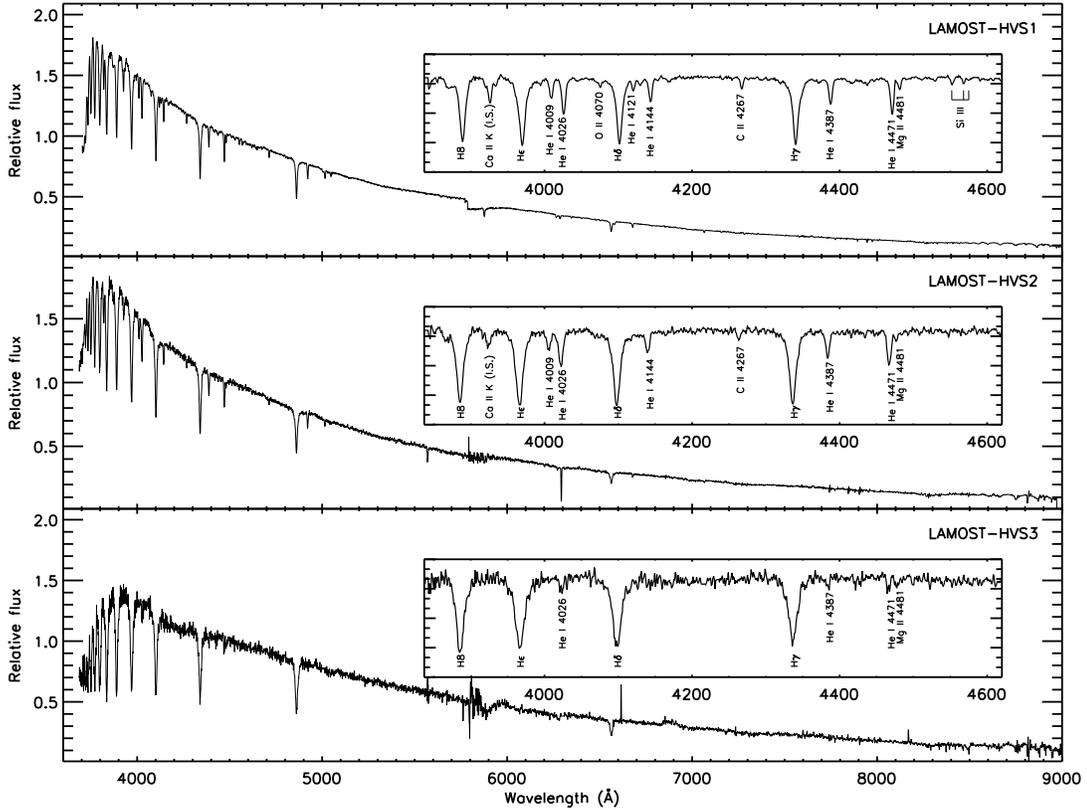}
\caption{LAMOST spectra of LAMOST-HVS1 (top), LAMOST-HVS2 (middle) and LAMOST-HVS3 (bottom).
The inset in each panel shows the enlarged normalized blue-arm spectrum} for a better view of the various spectral features detected.
\end{center}
\end{figure*}

\section{Data and target selection}
\subsection{Data}
LAMOST is a 4-metre, quasi-meridian reflecting Schmidt telescope equipped with 4000 fibers distributed in a field of view of $5^{\circ}$ in diameter (Cui et al. 2012).
It can simultaneously collect up to 4000 optical ($\lambda\lambda$ 3700 -- 9000), low resolution ($R$\,$\sim$\,$1800$) spectra per exposure.
After one year Pilot Surveys (Sep. 2011 -- Jun. 2012), the five-year LAMOST Regular Surveys were initiated in the fall of 2012 and completed this summer of 2017.
The scientific motivations and target selections of the surveys are described in Zhao et al. (2012), Deng et al. (2012) and Liu et al. (2014).
The Phase-II LAMOST Pilot Surveys are expected to start this September, and the Regular ones in September, 2018 and lasting for another five years.
With LAMOST Stellar Parameter Pipeline at Peking University (LSP3; Xiang et al. 2015, 2017) that we developed, we have derived line-of-sight velocities\footnote{We note that values of $v_{\rm los}$ yielded by LSP3 have been corrected for a systematic offset of 3.1 km\,$^{-1}$ (Xiang et al. 2015).} and basic atmospheric parameters (effective temperature $T_{\rm eff}$, surface gravity log\,$g$, metallicity [Fe/H]) from approximately 6.5 million stellar spectra (of 4.4 million unique stars) of SNRs higher than 10, collected with LAMOST by June 2016.
The above data set, to be publicly available soon as the third release of the LAMOST Spectroscopic Survey of the Galactic Anti-center (LSS-GAC) value-added catalogues (Huang et al. in preparation), is used for the purpose of the current study.

\subsection{Target selection}
With this tremendous data set, we first search for old, low-mass HVSs (i.e. of F/G/K spectral type or 4000\,$\le T_{\rm eff} \le$\,6500\,K).
This has been carried out by Huang et al. (2017, hereafter H17) in deriving the Galactic escape velocities at $r$ between 5 and 14\,kpc. 
From  over 3 million unique low-mass stars observed with LAMOST at distances 4-6\,kpc from the Sun, they find a total of 692 high radial velocity stars with $|v_{\rm rf}| \ge 300$\,km\,s$^{-1}$. 
The fastest of them has a Galactic rest-frame radial velocity of 465\,km\,s$^{-1}$, substantially lower than the local Galactic escape speed (e.g. Piffl et al. 2014; H17).
The fact that no radial-velocity low-mass HVS candidates have been found in the LAMOST spectroscopic surveys is consistent with earlier work using both SDSS and LAMOST data (e.g. Kollmeier et al. 2010; Li et al. 2012, 2015; Zhong et al. 2014).
 This fact places new constraints on the mechanism of low-mass HVS ejection as a function of mass, stellar population (e.g. metallicity, age, 
multiplicity) and velocity (e.g. Bromley et al. 2006; Kollmeier et al. 2010).
Our current results suggest that the initial mass function (IMF) of the parent population of HVS is probably top-heavy.
In H17, the identified high radial velocity stars (after excluding 165 disk stars) are used to derive the Galactic escape velocity curve for $r$ between 5 and 14\,kpc.
A Galactic mass model is further constructed from this curve combined with a prior on the local circular velocity.

We use the same data set to search for young, massive HVSs (i.e. of O/B/A spectral type or $T_{\rm eff} >$\,6500\,K).
We first select potential candidates by imposing a Galactic rest-frame radial velocity greater than 300\,km\,s$^{-1}$.
Here, the Galactic rest-frame radial velocity is defined as,
\begin{equation}
v_{\rm rf} = v_{\rm los}  + U_{\odot}\cos b \cos l + V_{\phi, \odot}\cos b \sin l + W_{\odot}\sin b\text{,}
\end{equation}
where  $V_{\phi, \odot}$, i.e. $V_{\rm c}$\,$(R_{0})$\,$+$\,$V_{\odot}$, is set to the value yielded by the proper motion of Sgr\,A$^{*}$ (Reid \& Brunthaler 2004) and $R_{0} = 8.34$\,kpc (Reid et al. 2014).
The values of the solar motion in the radial and vertical directions are adopted from Huang et al. (2015), i.e. ($U_{\odot}$,\,$W_{\odot}$)\,$=$\,($7.01$,\,$4.95$)\,km\,s$^{-1}$.
With the radial velocity cut, a total of 6602 spectra (of 6350 unique objects) of SNRs greater than 10 are selected.
We then visually check the spectra and find most of them are galaxies or AGNs mis-classified by the current pipeline.
With manual check and after excluding variables and M31/M33 objects, we are left with 126 high-mass HVS candidates.  

We then derive the distances of those HVS candidates.
For metal-rich ([Fe/H]\,$> -1$), (relatively) cool ($T_{\rm eff} \le10000$\,K) candidates, we adopt the absolute magnitudes yielded by LSP3.
For metal-poor ([Fe/H]\,$< -1$), (relatively) hot  ($T_{\rm eff} \ge10000$\,K) candidates, we apply the same Bayesian approach used by H17 in order to derive their absolute magnitudes.
For hot ($T_{\rm eff} > 10000$\,K) candidates, we first classify the stars with the code MKCLASS (Gray \& Corbally)\footnote{\url{http://www.appstate.edu/$\sim$grayro/mkclass/}} and then infer their absolute magnitudes accordingly (see Section\,3.1).
With the absolute magnitudes derived, we estimate the distances of the 126 HVS candidates from their apparent magnitudes (e.g. 2MASS $JHK_{s}$, APASS $BV$) after applying reddening corrections as given by the all-sky two-dimensional extinction map of Schlegel, Finkbeiner \& Davis (1998; hereafter SFD98) for stars at high latitudes ($|b| > 30^{\circ}$) or derived with the `star pair' technique (Yuan, Liu \& Xiang 2013) for stars close to the disk ($|b| \le 30^{\circ}$).
With the derived distances, we select real HVSs by requiring their Galactic rest-frame radial velocities greater than their corresponding Galactic escape velocities, as predicted by the Galactic mass model constructed by H17.
Finally, three HVSs are discovered, including the one found earlier by Z14.

\section{Results and discussion}
In this section, we present the properties of the three HVSs discovered in the LAMOST surveys, including the one by Z14, and discuss their origins.
The three HVSs have respectively Galactic coordinates of ($l$,\,$b$) = (221\fdg099543,\,35\fdg407234), ($l$,\,$b$) = (60\fdg398889, 45\fdg253192) and ($l$,\,$b$) = (165\fdg140088,\,-31\fdg200353), and are hereafter referred to LAMOST-HVS1, LAMOST-HVS2 and LAMOST-HVS3, respectively.

\subsection{Properties of the LAMOST HVSs}

\begin{figure}
\begin{center}
\includegraphics[scale=0.55,angle=0]{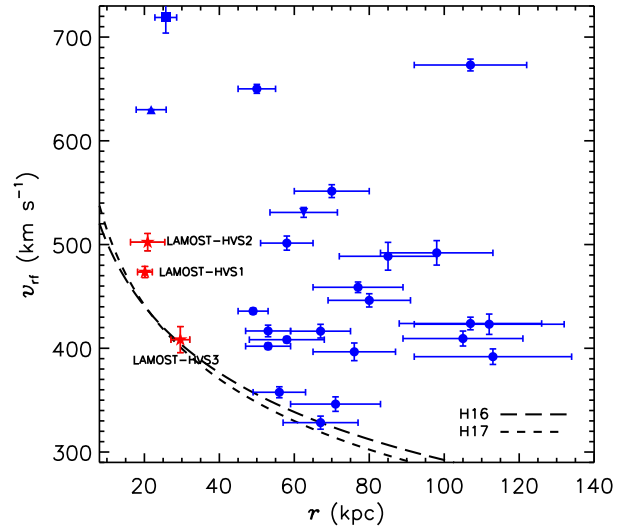}
\caption{Values of radial velocity in the Galactic rest-frame $v_{\rm rf}$ of 24 HVSs (blue circles; Brown et al. 2014; Brown 2015) discovered in other surveys and of LAMOST-HVS1,2,3 (red stars), plotted against the Galactocentric radius $r$.
The blue circles represent the 21 B-type HVSs discovered from systematic searches (e.g. Brown et al. 2014) while the blue box, triangle and inverted triangle represent the three serendipitously discovered HVSs: US\,708 (Hirsch et al. 2005), HD\,271791 and HE\,0437-5439 respectively.
We note that the velocity plotted of star HD\,271791 is the total velocity rather than the Galactic rest-frame radial velocity.
Black long and short dashed lines represent the Galactic escape velocity curve predicted by the Milky Way mass model constructed by Huang et al. 2016 and 2017, respectively.}
\end{center}
\end{figure}

\textbf{LAMOST-HVS1:}  As mentioned above, LAMOST-HVS1 is first discovered by Z14.
Its basic properties are given in Table\,1.
Hitherto, it has been targeted three times by LAMOST and the combined spectrum is presented in Fig.\,1.
From the combined spectrum, we derive a heliocentric radial velocity of $611.65 \pm 4.63$\,km\,s$^{-1}$, consistent with the value of Z14 within the uncertainties.
This corresponds to a Galactic rest-frame radial velocity of $473.52 \pm 5.35$\,km\,s$^{-1}$.
According to the spectral classification of MKCLASS, LAMOST-HVS1 is a B1IV/V star, again compatible with the result obtained by Z14.
Given the spectral type, we infer the absolute magnitude, mass, luminosity, $T_{\rm eff}$ and life time  of LAMOST-HVS1 using the look-up table provided by Mamajek\footnote{\url{http://www.pas.rochester.edu/$\sim$emamajek/EEM\_dwarf\_UBVIJHK\_colors\newline\_Teff.txt}}.
This gives a distance of LAMOST-HVS1 of  $13.91 \pm 1.97$\,kpc by combining the derived absolute magnitudes and the apparent magnitudes and the SFD98  extinction value.
By adopting a value of 8.34\,kpc for the Galactocentric distance of the Sun, $R_{0}$, the LAMOST-HVS1 is then found to have a Galactocentric radius  $r = 20.14 \pm 1.98$\,kpc at a height $Z = 8.06 \pm 1.14$\,kpc above Galactic plane.
Again, the estimates are in excellent agreement with the results of Z14.

\textbf{LAMOST-HVS2:}
The second HVS discovered by LAMOST is a B2V star (as given by MKCLASS) with mass of around 7\,$M_{\odot}$.
From the spectrum (see Fig.\,1) combining two observations by LAMOST, we measure a heliocentric radial velocity of $341.10 \pm 7.79$\,km\,s$^{-1}$, corresponding to a Galactic rest-frame radial velocity of $502.33 \pm 8.37$\,km\,s$^{-1}$.
Following the same approach described above, we infer the basic physical properties of LAMOST-HVS2 as listed in Table\,1, and further derive a distance of $22.24 \pm 4.57$\,kpc for it.
The estimated distance places  LAMOST-HVS2 at $r = 20.86 \pm 4.57$\,kpc and $Z = 15.80 \pm 3.25$\,kpc.
The Galactocentric radius of LAMOST-HVS2 is very close to that of LAMOST-HVS1.

\textbf{LAMOST-HVS3:}
LAMOST-HVS3 has hitherto been observed by LAMOST only once and the spectrum is presented in Fig.\,1.
Its heliocentric radial velocity is measured to be $361.38 \pm 12.52$\,km\,s$^{-1}$, corresponding to a velocity of $408.33 \pm 12.57$\,km\,s$^{-1}$ in the Galactic rest-frame.
MKCLASS classifies LAMOST-HVS3 as a B7V star.
We again infer the basic physical properties of LAMOST-HVS3 (see Table\,1) according to its spectral classification and further estimate its distance.
The estimated mass is about 4\,$M_{\odot}$ with a $T_{\rm eff} = 14\,000$\,K.
The distance of LAMOST-HVS3 is $22.32 \pm 2.50$\,kpc and this distance places it at $r = 29.59 \pm 2.51$\,kpc from the GC and at $Z = -11.56 \pm 1.30$\,kpc below the Galactic plane.

In Fig.\,2, we plot the radial velocity in the Galactic rest frame $v_{\rm rf}$ as a function of Galactic radius $r$ for all the known HVSs, including the three HVSs discovered with LAMOST.
As the figure shows, the three HVSs discovered by LAMOST, together with two other HVSs found by other groups, are the five closest HVSs.
They are all within 30\,kpc of the GC.

\begin{figure}
\begin{center}
\includegraphics[scale=0.55,angle=0]{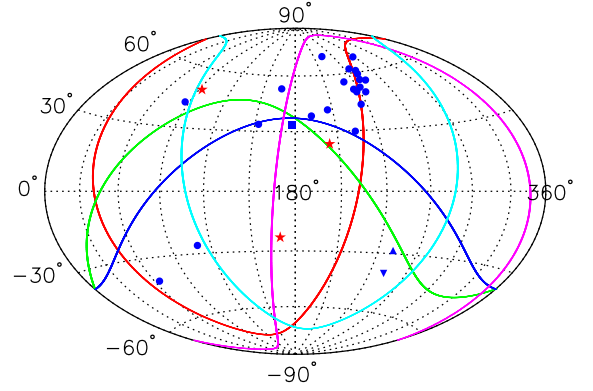}
\caption{Galactocentric view of the spatial distribution of confirmed HVSs (including the three discovered with LAMOST) in the Galactic coordinate system.
The definitions of the different symbols are same as in the Fig.\,2.
The great circles represent the planes (projected to infinity) of the various young stellar structures near the GC, including the CWS (red), the CCWS (cyan), the outer CWS (green), the Narm (blue) and the bar (magenta).}
\end{center}
\end{figure}

\subsection{Origins of the LAMOST HVSs}
In this section, we discuss the possible origins of the three HVSs discovered with LAMOST.

First, we try to constrain their origins by their spatial distribution.
According to Lu et al. 2010 and Zhang et al. (2010, 2013), for HVSs originating from the GC, their spatial distribution can be used to trace the parent population of their progenitors directly.
Lu et al. (2010) suggest that HVSs identified hitherto may arise from the two young star-forming regions near the GC (see Fig.\,3): a clockwise young stellar disk (CWS) and a northern arm (Narm).
Similar to Lu et al. (2010), we plot in Fig.\,3 the spatial distribution in the Galactic coordinate (system as viewed from the GC) of the three HVSs discovered by LAMOST.
The planes (projected to infinity) of several young stellar structures near the GC are also overplotted, including the CWS, the Narm, the counter-clockwise young stellar disk (CCWS), the outer warped CWS and the bar.
As pointed out by Z14, LAMOST-HVS1 is spatially very close to the outer CWS but also not far from the Narm.
LAMOST-HVS2 is spatially very close to the CCWS but also not far from  the outer CWS.
LAMOST-HVS3 is very close to the bar.
The spatial association may suggest that all  the three LAMOST HVSs may originate from the GC and their progenitors are spatially associated with young stellar structures near the GC, including the outer CWS, the CCWS, and the bar.
In addition, the discovery of the three LAMOST HVSs reduces the apparent anisotropy of spatial distribution of known HVSs on the sky.

Secondly, assuming that the three LAMOST HVSs arise from the GC, one can calculate their flight times. 
It should be noted that the flight times derived here are actually upper limits since it is assumed that the observed radial velocities represent the full space motion of the HVSs.
 The results are listed in Table\,1.
 As one can see, the flight times of LAMOST-HVS1 and 2 are larger than their life times, implying that they do not have enough time to travel to their current positions from the GC.
 If their GC origin holds, the possible explanation is that these two stars are blue stragglers that have experienced a similar process to HVS HE\,0437-5439.
 The flight time ($\sim$\,58\,Myr) of LAMOST-HVS3 is substantially smaller than its life time ($\sim$\,80\,Myr), suggesting that it has enough time to travel to its current position even there is a few million delay in its ejection from the GC (Brown et al. 2012).
 
 Finally, accurate proper motion measurements are required in order to better constrain the origin of the three HVSs discovered with LAMOST.
The current measurements have uncertainties (systematic plus random) too large to make a conclusive analysis [see similar discussions in Z14 and Kenyon et al. (2014) for LAMOST HVS1].
Fortunately, all the three stars are quite bright and the upcoming {\it Gaia} data release should solve this problem.
 
\section{Summary}
We have carried out a  systematic search for HVSs in nearly 6.5 million qualified (SNRs\,$> 10$) stellar spectra of 4.4 unique stars collected upto June 2016 by LAMOST spectroscopic surveys, and have discovered three HVSs, including the one found earlier by Z14.
The two newly discovered HVSs are respectively a B2V type star of $\sim$\,7\,M$_{\rm \odot}$ with a Galactic rest-frame radial velocity of 502\,km\,s$^{-1}$ at a Galactocentric radius of $\sim$\,21\,kpc and a B7V type star of $\sim$\,$4$\,M$_{\rm \odot}$  with a Galactic rest-frame radial velocity of 408\,km\,s$^{-1}$ at a Galactocentric radius of $\sim$\,30\,kpc.

We discuss the possible origins of the three HVSs based on their spatial positions and flight times.
The three HVSs are all spatially associated with known young stellar structures near the GC, which supports a GC origin for them.
However, two of them, i.e. LAMOST-HVS1 and 2, have life times smaller than their flight times, indicating that they do not have enough time to travel from the GC to the current positions unless they are  blue stragglers (as in the case of HVS HE 0437-5439).
The third one (LAMOST-HVS3) has a life time larger than its flight time and thus does not have this problem.
The upcoming accurate proper motion measurements by {\it Gaia} should provide a direct constraint on their origins.        

Finally,  we expect more HVSs to be discovered by the ongoing LAMOST spectroscopic surveys and thus to provide further constraint on the nature and ejection mechanisms of HVSs.

 \section*{Acknowledgements} 
 We would like to thank the referee for his/her helpful comments.
 This work is supported by the National Key Basic Research Program of China 2014CB845700, the China Postdoctoral Science Foundation 2016M600849 and the National Natural Science Foundation of China 11473001 and 11233004.  
The LAMOST FELLOWSHIP is supported by Special fund for Advanced Users, budgeted and administrated by Center for Astronomical Mega-Science, Chinese Academy of Sciences (CAMS). 
We thank Youju Lu and Fupeng Zhang for valuable discussions.
 
The Guoshoujing Telescope (the Large Sky Area Multi-Object Fiber Spectroscopic Telescope, LAMOST) is a National Major Scientific Project built by the Chinese Academy of Sciences. Funding for the project has been provided by the National Development and Reform Commission. LAMOST is operated and managed by the National Astronomical Observatories, Chinese Academy of Sciences.

\end{document}